\documentclass[traditabstract]{aa} 
\usepackage{graphicx}
\usepackage{amsmath}
\usepackage[USenglish]{isodate}
\usepackage{natbib,twoopt}
\usepackage{hyperref} 
\usepackage{color}
\bibpunct{(}{)}{;}{a}{}{,}

\newcommand{\srs}{slow--rotator sequence}

\defcitealias{LS15}{LS15}
\defcitealias{SL20}{SL20}

\begin{document}

\title{Rotational evolution of slow--rotator sequence stars.}
\subtitle{II. Modeling the wind braking and the rotational coupling \\ in the entire mass range of solar-like stars.}

\titlerunning{Mass-dependent wind braking and rotational coupling}

\author{
F. Spada\inst{1}
\and
A.~C. Lanzafame\inst{1,2}
}

\authorrunning{F. Spada \& A.~C. Lanzafame}

\offprints{F. Spada}

\institute{
Universit\`a di Catania, Dipartimento di Fisica e Astronomia, Sezione Astrofisica, Via S. Sofia 78, 95123 Catania, Italy 
\and
INAF-Osservatorio Astrofisico di Catania, Via S. Sofia 78, 95123 Catania, Italy
}

\date{}

\abstract{
In recent years, ground- and space-based photometric surveys have characterized the rotational evolution of solar-like stars to an unprecedented level of detail.
In this work we focus on the slow--rotator sequence, an emergent feature recognizable in the color--period diagram of Galactic open clusters. 
Understanding the evolution of this sequence is a promising avenue to formulate a robust rotation period--mass--age relation, which can be used to estimate stellar ages.
Our model of the rotational evolution of stars on the slow-rotator sequence takes into account magnetized wind braking and the rotational decoupling between the radiative interior and the convective envelope. 
This decoupling naturally develops as the internal redistribution of angular momentum lags behind the loss of angular momentum at the stellar surface, and is parameterized in the model by a rotational coupling timescale.
Using literature data on rotation and membership of stars in a selection of open clusters of age between $100$ Myr and $4$ Gyr, we constrain the mass dependence of the two competing processes of wind braking at the surface and angular momentum transport in the interior.
Consistently with our previous findings, our best-fitting model requires a mass-dependent coupling timescale; this result is insensitive to the details of the wind braking model used.
We show that the mass dependence of the coupling timescale follows a broken power-law in the entire solar-like mass range ($0.4$--$1.25\, M_\odot$), with the exponent change occurring at $\approx 0.6 \, M_\odot$.
At the same time, our approach can be used to infer semi-empirically the mass dependence of the wind braking model that best fits the observational constraints.
Based on our findings, we propose a novel wind braking law with a particularly simple mass term, directly proportional to the moment of inertia of the convective envelope of the star.
}

\keywords{Stars: rotation -- Stars: evolution -- Stars: late-type -- open clusters and associations: general}

\maketitle

\section{Introduction}
\label{intro}

In stars with outer convective envelopes, mass loss from a magnetized stellar wind, powered by dynamo action, results in a braking torque that drives a significant rotational evolution. 
The surface rotation period of a $1\, M_\odot$ star, for instance, decreases by a factor of $\approx25$ over the approximately $5$ Gyr interval between zero-age main sequence and solar age \citep{Barnes_ea:2016}. 
In a similar fashion, stars with masses between $0.4$ and $1.25\, M_\odot$ undergo a ``solar-like'' rotational evolution.

The most stringent observational constraints on solar-like rotational evolution come from surface rotation periods measured via photometric monitoring for stars members of coeval populations of known age, such as Galactic open clusters.
Observational color--period diagrams of open clusters, in particular, provide snapshots of the rotational state of stars of different mass, inferred from the observational proxy of their photometric color, at approximately the same age, which can be independently and reliably estimated from isochrone fitting \citep[e.g.,][]{Demarque_Larson:1964}. 

In a seminal paper, \citep{Barnes:2003} pointed out that a sequence of comparatively slow rotators gradually emerges in the color--period diagram of clusters of age between $30$ and $600$ Myr, and that the timescale for transition from fast- to slow-rotator status is mass-dependent, i.e., solar-type stars are already on the sequence by $\approx 100$ Myr, while stars of later spectral type converge onto it at progressively later ages.
Although the transition from fast to slow rotator is still not fully understood (see, e.g., \citealt{Barnes:2010}, \citealt{Matt_ea:2012}, \citealt{Brown:2014}, \citealt{Lanzafame_ea:2019}), the details of the underlying process are mostly forgotten after the first $\approx 100$ Myr.
The subsequent evolution of the \srs{} is remarkably regular, and, somewhat analogously to the main sequence in the classical color--magnitude diagram, encodes key information on the solar-like rotational evolution process as a whole.

The emergent properties of the \srs{} (i.e., its overall evolution and its mass dependence) place strong constraints on two key physical processes, whose theoretical understanding remains elusive, namely, the removal of angular momentum from the surface mediated by the magnetized stellar wind, and the transport of angular momentum in the stellar interior.
The former, usually referred to as (magnetized) wind braking, is the main driver of the rotational evolution of solar-like stars during their main sequence lifetime, while the latter, remarkably, has second-order, yet observable consequences, and thus provides a window into the physics of stellar interiors.

Wind braking in solar-like stars has been known observationally and, at least in broad terms, understood theoretically, for more than fifty years (\citealt{Kraft:1967}, \citealt{Skumanich:1972};\citealt{Schatzman:1962}, \citealt{Weber_Davis:1967}, \citealt{Kawaler:1988}).
The general picture established in these early works is still accepted today: the wind effectively applies on the star a braking torque, which is proportional to the mass loss rate and to the surface rotation rate, and, crucially, is greatly amplified by the fact that the wind outflows are dominated by the dynamo magnetic fields up to several stellar radii from the surface.
This basic mechanism is responsible for the observed spin-down of cool stars, which, to the lowest order of approximation, is captured by the \citealt{Skumanich:1972} law: $P_{\rm rot} \propto \sqrt{\rm age}$.

Large rotation period surveys (Kepler, K2, TESS), together with Gaia-informed cluster ages, have produced dense rotation--age--mass data sets from tens of Myr to several Gyr \citep[e.g.,][and references therein]{Van-Lane_ea:2025}, revealing a rich set of features that simple spin-down relations, such as the Skumanich law, cannot fit. 
For instance, the spin-down mass dependence cannot be fully factored out from the age dependence \citep{Meibom_ea:2009, Meibom_ea:2011, LS15}, it may anomalously slow down (``stall'') at certain ages \citep{Curtis_ea:2020}, and it may even shut down beyond solar age  \citep{vanSaders_ea:2016, Metcalfe_ea:2025}.
The torque normalization and its scaling relations with stellar rotation, mass, and other physical parameters required to reproduce the observed features of rotational evolution are the topic of active research, and several wind braking laws have been proposed in the literature \citep[e.g.,][]{Krishnamurthi_ea:1997, Matt_ea:2012, Reiners_Mohanty:2012, Matt_ea:2015, Garraffo_ea:2018, SL20}.

The measurement of the rotation profile of the solar interior via helioseismic inversion \citep{CD_Schou:1988, Schou_ea:1998}, on the other hand, indirectly constrains the transport of angular momentum in the interior of solar-like stars \citep[e.g.,][]{Eggenberger_ea:2005}.
In particular, the nearly uniform rotation of the solar interior implies that efficient angular momentum redistribution must occur during the main sequence.
An independent confirmation comes from asteroseismic inversions in subgiants and red giant stars, which reveal that the cores of these stars rotate faster than the envelopes, but much more slowly than predicted by models without efficient angular momentum transport, again implying a strong coupling between core and envelope during evolution \citep[e.g.,][]{Mosser_ea:2012, Cantiello_ea:2014, Deheuvels_ea:2014, Fuller_ea:2014, Spada_ea:2016, Eggenberger_ea:2019, Deheuvels_ea:2020, Moyano_ea:2023}.

Standard models with only hydrodynamic processes fail to match asteroseismic constraints \citep[e.g.,][]{Pinsonneault_ea:1989, Eggenberger_ea:2005}.
Including magnetic fields or internal gravity waves improves the agreement, but the exact combination and efficiency of these mechanisms remain uncertain \citep{Charbonneau_MacGregor:1993, Ruediger_Kitchatinov:1996, Charbonnel_Talon:2005, Oglethorpe_Garaud:2013, Aerts_ea:2019}.

In summary, in spite of considerable theoretical and observational progress, a comprehensive physical picture of either wind braking or rotational coupling in solar-like stars is still lacking.
Building on previous work, in this paper we focus in particular on improving the mass-scaling of these two processes, based on the constraints that can be extracted from the observed \srs{}.

A key feature of our modeling paradigm is a two-zone description of the rotation profile in the interior of the star \citep{MacGregor_Brenner:1991}.
This simple, but versatile approach introduces a differential rotation between the radiative interior and the convective envelope, parametrizing the angular momentum transport with a rotational coupling timescale \citep[e.g.,][]{Spada_ea:2011, Gallet_Bouvier:2013, Gallet_Bouvier:2015}).
A short coupling timescale keeps a star rotating as a solid body essentially at all times, while a longer timescale allows for some angular momentum to be temporarily stored in the radiative interior, to resurface later and temporarily offset the evolution toward longer rotation periods.
Such non-monotonic rotational evolution of the surface rotation periods suffices to capture quantitatively the observed stalled magnetic braking \citep{Curtis_ea:2019, Curtis_ea:2020, Angus_ea:2020, Gordon_ea:2021, Santos_ea:2025}.

In this work we derive the parametric dependence of the coupling timescale on stellar mass across the full range of stars with solar-like interior structure, confirming and extending the results of \citet{LS15} and \citet{SL20} (hereafter \citetalias{LS15} and \citetalias{SL20}, respectively).
We find that this dependence is robust to different choices of the wind braking law.
We further derive a semi-empirical mass term for the wind braking law and show that its functional behavior is well described by a scaling with a single stellar parameter, namely, the moment of inertia of the convective envelope.

This paper is organized as follows:
In Section~\ref{data} we discuss the criteria used to construct the data set which we use to constrain our models; in Section~\ref{methods} we describe the data preparation and treatment, the formulation of our rotational evolution model, and our fitting procedure; our results are presented in Section~\ref{results}, and discussed in Section~\ref{discussion}; we summarize our conclusions in Section~\ref{conclusions}.

\begin{table}[htp]
\caption{Parameters of the open clusters used in this work. Rotation periods and membership are from the main reference.}
\begin{center}
\begin{tabular}{cccc}
\hline
\hline
Name              & Main ref. & $E(B-V)$  & Age (Gyr) \\
\hline
Pleiades          & (1)                     &  0.045 (6)   & 0.12  (9)       \\
NGC 3532       & (2)                     &  0.022 (6)   & 0.30  (11)     \\
M37                 & (3)                     & 0.246  (7)   & 0.50  (12)    \\
Praesepe         & (1)                    & 0.027   (6)   & 0.70  (9)     \\
NGC 6811        & (1)                    &  0.047  (7)   & 1.00  (9)    \\
NGC 6819       &  (1)                    &  0.1      (8)  & 2.50   (9)  \\
Ruprecht 147  &  (1)                    &  0.1      (9)  & 2.70   (9)    \\
M67                 & (4), (5)              &  0.04    (10) & 4.00  (13)    \\
\hline
\end{tabular}
\end{center}
(1): \citet{Curtis_ea:2020}; (2): \citet{Fritzewski_ea:2021}; (3): \citet{Hartman_ea:2009}; (4): \citet{Dungee_ea:2022}; (5): \citet{Gruner_ea:2023};
(6): \citet{GCB:2018}; (7): \citet{Godoy-Rivera_ea:2021}; (8): \citet{Bragaglia_ea:2001}; (9): \citet{Curtis_ea:2020}; (10): \citep{Taylor:2007};
(11): \citet{Fritzewski_ea:2021}; (12): \citet{Hartman_ea:2009}; (13): \citet{Gruner_ea:2023}.
\label{table_oc}
\end{table}

\section{Data}
\label{data}

Following up on our previous work \citepalias{LS15, SL20}, we extend our modeling of the \srs{} to the widest possible domain of applicability.
To this end, we fit our models with rotation period data of stars in the mass range $0.4$--$1.25 \, M_\odot$, and in a (nominal) age range of $0.1$--$4$ Gyr.
By definition, the two-zone model approach (see Section~\ref{twozonemodel}) is applicable to stars having a solar-like interior structure during the main sequence, namely, a sufficiently developed radiative core (which implies the lower limit $M_* \gtrsim 0.35 \, M_\odot$) and a sufficiently massive convective envelope (implying $M_* \lesssim 1.3\, M_\odot$).
The age range considered, on the other hand, is limited by the occurrence of the \srs{} in open clusters, as well as the availability of observational data. 
The youngest cluster in which the \srs{} is significantly developed is the $\approx 100$ Myr old Pleiades, while the oldest open cluster for which rotational information is currently available is the $\approx 4$ Gyr old M67). 
Further restrictions on the applicability of our models may arise from the so-called ``weakened magnetic braking'' phenomenon \citep{vanSaders_ea:2016}.
We do not attempt to model the influence of metallicity on the rotational evolution of solar-like stars \citep[e.g.,][]{Amard_Matt:2020}.

Our dataset consists of a compilation of rotation periods, together with membership and photometric information, for stars that are members of open clusters in which the \srs{} (also known as the ``I-sequence'', see \citealt{Barnes:2003}) is readily identifiable and well constrained by the available data.
The open clusters selected for this work are listed in Table~\ref{table_oc}, which also includes nominal ages and visual reddening, with their respective references. 

Our selection was based on three main criteria: i) existence and adequate sampling of the \srs{}; ii) dense logarithmically spaced coverage of the $0.1$-$4$ Gyr time span; iii) richness and quality of the available data (e.g., accurate rotation periods from space-based photometric surveys, reliable membership information).
Our selection of open clusters is clearly not the only possible one, nor it is meant to be exhaustive or fully representative of the entire rotational dataset available in the literature to-date.

We adopted the de-reddened Gaia $G_{\rm BP}-G_{\rm RP}$ color as our observational proxy of stellar mass. 
We used Gaia colors from the latest data release \citep[DR3;][]{Andrae_ea:2023, DeAngeli_ea:2023} for the stars in each cluster. 
We applied the de-reddening procedure described in \citet{GCB:2018}, with the polynomial fit coefficients reported in that paper. 

In this work, we do not attempt to address directly the uncertainty on the age or on the reddening.
In particular, we chose not to include the ages of the clusters as free parameters in our fit.
The uncertainty of ages is therefore not explicitly taken into account in our fit, but we have verified that age variations within a reasonable range ($\approx 10\%$) does not affect our results significantly.
The robustness of our models to changes in the adopted nominal ages of the clusters will be further commented on in Section~\ref{chronoflow}.

\section{Methods}
\label{methods}

Our procedure to model the \srs{} consists of two steps.
First, for each of the open clusters in our catalog (see Section \ref{data}), we identify the stars belonging to the \srs{}, and from this we determine the empirical color--period relation of the sequence of the cluster by means of a non-parametric fit.
After transforming the color into mass by isochrone interpolation, we obtain the empirical mass--period relation of the \srs{} as a function of age between $0.1$ and $4.0$ Gyr for the mass range $0.4$--$1.25 \, M_\odot$.
In the second step, we used these empirical constraints to fit the parameters of our rotational evolution model. 
This second fit is based on a standard least-squares approach. 
The rest of this Section provides a detailed description of our methodology.

 \begin{figure}
\begin{center}
\includegraphics[width=0.49\textwidth]{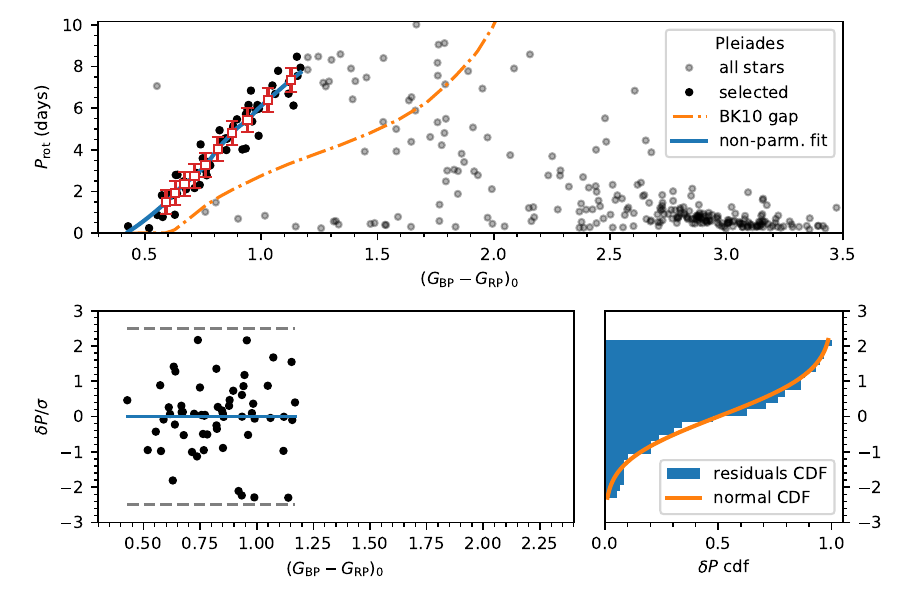}
\caption{Non--parametric fitting of the \srs{} of the Pleiades. Top panel: color--period diagram; the empirical constraints on the \srs{} extracted from the non-parametric fit are shown as empty red squares with error bars. 
Bottom left: normalized fit residuals;  the grey dashed lines represent deviations of $\pm 2.5 \times \sigma$ from the mean. Bottom right: cumulative distribution of the normalized fit residuals compared with that of a normal distribution.}
\label{fig:ple_npf}
\includegraphics[width=0.49\textwidth]{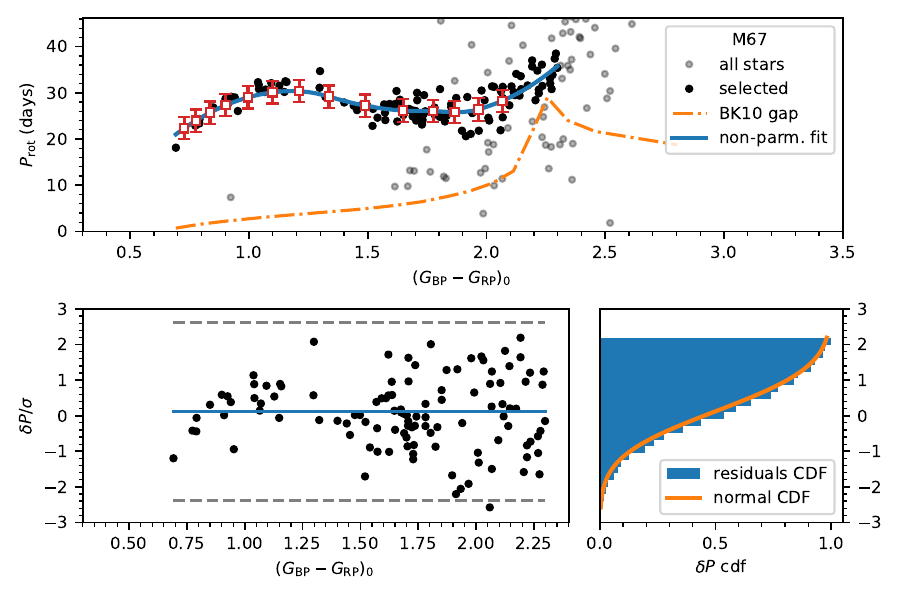}
\caption{Same as Figure~\ref{fig:ple_npf}, but for M67.}
\label{fig:m67_npf}
\end{center}
\end{figure}

\subsection{Selection of the \srs{}}
\label{nonparmfit}

To identify the stars belonging to the \srs{} in a given cluster we adopted the same iterative procedure used in \citetalias{LS15}.
Starting from an initial selection of candidate members, we obtain an approximate color--period relation on the sequence via a non--parametric fit. 
The rotation period residuals of the candidate members with respect to the current fit are then evaluated, and stars with residuals larger than $2.5$ standard deviations are discarded.
This process is repeated until convergence ($\lesssim 5$ iteration are usually sufficient). 
The standard deviation of the residuals with respect to the converged fit is taken to be representative of the spread of the \srs{} of that particular cluster. 
We adopt this converged standard deviation as a measure of the uncertainty in the empirical color--period distribution. 
Note that both intrinsic spread, astrophysical in origin, and spread due to observational errors, are likely to contribute to this uncertainty.
Distinguishing the exact source of this spread is not necessary for our purposes, as explained below.

The candidate stars included in the starting guess are those having rotation period above the empirical ``gap curve'' of \citet{Barnes_Kim:2010} in the color--period diagram.
The non-parametric fit is based on the local polynomial regression fit procedure (LOESS; \citealt{Cleveland_ea:2017}), as implemented in the statistical package \texttt{R}. 
The smoothing parameter of the LOESS fit was adjusted by trial and error, and the final values are between $0.5$ and $0.8$ for all clusters.
In general, we only considered stars with $(G_{\rm BP}-G_{\rm RP})_0 \leq 2.3$, i.e., bluer than the transition to fully convective interiors (except for the Pleiades, see below).
Although, as discussed at length by \citetalias{LS15}, there is no strong theoretical motivation to expect that the stars on the \srs{} should be distributed normally around its middle ridge, this procedure is sufficiently robust and accurate to select the sequence members, and provides an internally consistent approach to construct a fit of the sequence for each cluster in our sample.

In practice, the identification of the sequence is almost trivial for the clusters having very few or no fast rotators in the $(G_{\rm BP}-G_{\rm RP})_0$ range considered. 
This is the case for NGC~3532, M37, Praesepe, NGC~6811, NGC~6819, Ruprecht~147. 
 For these clusters, the selection procedure described above is mostly required to address the slow--rotator status in specific color subranges, and to obtain a fit of the \srs{}, with its associated uncertainty, in a way that is as uniform as possible across the clusters. 
 The Pleiades and M67, on the other hand, deserve some additional comments. 
 
The Pleiades cluster is sufficiently young to have a significant population of fast rotators and ``gap'' stars into which the \srs{} merges, and becomes hard to disentangle from, at $(G_{\rm BP}-G_{\rm RP})_0 \gtrsim 1.2$ (see Figure~\ref{fig:ple_npf}).
We therefore introduce a cutoff at this color, to ensure that our models are fitted only to bona-fide slow rotators.
A similar situation arises for the late K-/early M-type stars in M67, since these stars transition to the \srs{} with a timescale of the order of a few Gyrs (see Figure~\ref{fig:m67_npf}).
In this case, however, we obtained a satisfactory selection of the \srs{} using the iterative procedure based on the non--parametric fit described above, albeit with a converged $\sigma$ value that is significantly larger than for the other clusters.

Upon convergence, our iterative procedure provides a non-parametric fit and a standard deviation of its residuals which we adopt as the empirical color--period distribution of the \srs{} and its uncertainty, respectively, at the nominal age of that cluster.
The empirical relations are finally converted from color--period to mass--period by interpolation, using isochrones of solar metallicity and appropriate age extracted from the MIST database \citep{Choi_ea:2016}.

\subsection{The rotational evolution model}
\label{twozonemodel}

Our rotational evolution model follows the two-zone paradigm \citepalias[see also][and references therein]{LS15, SL20}.
The key assumption of the two-zone model is that, while the radiative interior and the convective envelope of a solar-like star rotate with approximately uniform velocity, a significant rotational gradient can develop at the interface between these two regions \citep{MacGregor_Brenner:1991}.
In other words, we assume that the rotation profile of a main-sequence, solar-like star at any time $t$ can be approximated with a step function, with the transition between the angular velocity of the radiative zone, $\Omega_c(t)$, to that of the convective envelope, $\Omega_e(t)$, located at the bottom of the convective envelope.

The model describes the evolution of $\Omega_c$ and $\Omega_e$ taking into account the following physical processes:
\begin{itemize} 
\item The initial conditions at $t=t_0 \lesssim$ 1 Myr assume a fully convective, uniformly rotating star with initial period $P_0$, i.e., $\Omega_c(t_0)=\Omega_e(t_0) = 2\pi/P_0$; as soon as a radiative core starts to form (within a few to tens of Myr since the birth line, depending on the mass of the star), $\Omega_c$ and $\Omega_e$ can assume distinct values. 
\item According to the disk-locking paradigm \citep{Koenigl:1991}, we assume that the magnetic interaction of the young star with its circumstellar disk keeps the angular velocity of the convective envelope constant for the lifetime of the disk ($\tau_{\rm disk} \lesssim 5$ Myr, \citealt{Hernandez_ea:2008}).
\item As the star evolves, a radiative core forms during the pre-main sequence within an initially fully convective structure, and subsequently evolves into the fully developed radiative interior region around the ZAMS epoch; changes in the stellar radius, the relative size and mass of the radiative zone and convective envelope, and their moments of inertia are taken into account, based on non-rotational stellar evolution tracks (see below for details).
\item The magnetized stellar wind effectively applies a braking torque at the surface \citep[e.g.,][]{Schatzman:1962, Kraft:1967, Skumanich:1972}, removing angular momentum from the convective envelope at the rate $\left.\frac{dJ}{dt}\right|_{\rm wb}$; the choice of this ``wind braking law'', which is one ofs the largest sources of uncertainty in our model, is further discussed below.
\item As the convective envelope loses angular momentum through the magnetized wind, an angular momentum excess develops between the radiative interior and the convective envelope; this excess is redistributed over time, as evidenced by the almost uniform solar rotation profile \citep{Schou_ea:1998}, but this process is still not well understood, so we model it parametrically by introducing a rotational coupling timescale $\tau_{\rm cpl}$.
\end{itemize}

Mathematically, the model consists of two coupled ordinary differential equations for the evolution of $\Omega_c$ and $\Omega_e$:
\begin{align}
\nonumber
& t  \leq \tau_{\rm disk}: \\
\nonumber
& I_c \, \dfrac{d\Omega_c}{dt} = + \dfrac{2}{3} \dfrac{dM_c}{dt} R_c^2 \,\Omega_e - \dfrac{\Delta J}{\tau_{\rm cpl}}  - \dfrac{dI_c}{dt} \Omega_c;
\\
\nonumber
& \phantom{I_e} \, \dfrac{d\Omega_e}{dt} = 0;
\\
\label{tzm}
\\
\nonumber
& t  > \tau_{\rm disk}: \\
\nonumber
& I_c \, \dfrac{d\Omega_c}{dt} = + \dfrac{2}{3} \dfrac{dM_c}{dt} R_c^2 \,\Omega_e - \dfrac{\Delta J}{\tau_{\rm cpl}}  - \dfrac{dI_c}{dt} \Omega_c;
\\
\nonumber
& I_e\, \dfrac{d\Omega_e}{dt} = - \dfrac{2}{3} \dfrac{dM_c}{dt} R_c^2 \, \Omega_e + \dfrac{\Delta J}{\tau_{\rm cpl}}   - \dfrac{dI_e}{dt} \Omega_e + \left.\frac{dJ}{dt}\right|_{\rm wb},
\end{align}
where $M_c$, $R_c$, $I_c$ are the mass, radius, and moment of inertia of the radiative interior, $I_e$ is the moment of inertia of the convective envelope, and $\Delta J = \frac{I_c\, I_e}{I_c+I_e} (\Omega_c - \Omega_e)$ (cf. \citealt{MacGregor_Brenner:1991}).

The equations are integrated using the ODE solver \texttt{solve\_ivp}, which is part of the \texttt{scipy.integrate} package.
We selected the integration method `RK45', an embedded explicit Runge-Kutta method of order 5(4).
During the course of the integration, the stellar structure parameters are obtained by interpolation from stellar evolutionary tracks, using \texttt{splrep} with pre-built spline fits constructed with \texttt{BSpline}, both part of \texttt{scipy.interpolate}. 
The evolutionary tracks were constructed with solar metallicity and mass in the range $0.4$--$1.25\, M_\odot$ in steps of $0.05\, M_\odot$ using the YREC stellar evolution code in its non-rotational configuration \citep{Demarque_ea:2008}.

Our wind braking law follows the classic structure:
\begin{equation}
\label{wind_braking}
\left.\frac{dJ}{dt}\right|_{\rm wb} = - K_w \cdot f_{M_*} \cdot \left(\frac{\Omega_e}{\Omega_\odot}\right)^3,
\end{equation}
where $K_w$ is a scaling constant (in the following, numerical values of $K_w$ are quoted in units of $1.13 \cdot 10^{47}$ g cm$^2$ s for convenience, cf. \citealt{Kawaler:1988}). 
The $\Omega_e^3$ dependence is based on theoretical and empirical estimates of the scaling of the surface magnetic field with surface rotation \citep[``dynamo law''; e.g.,][]{Stix:1972, Linsky_Saar:1987, Kawaler:1988}, leading to a rotational evolution approximately following the $P_{\rm rot} \propto t^{1/2}$ Skumanich law \citep{Skumanich:1972}.
We do not attempt to introduce any saturation effect of the dynamo law, since the present work focuses on the evolution of the \srs{}, which is composed of stars in the non-saturated regime (by definition).

The term $f_{M_*}$, on the other hand, imparts a mass dependence to the wind braking law by specifying an explicit dependence on selected stellar structure quantities (note that in this way an additional indirect, mostly negligible time dependence, due to the evolution of the stellar parameters, is also introduced). 
This term is crucial to model satisfactorily the \srs{} (see \citealt{Barnes_Kim:2010}, \citetalias{LS15, SL20} for details).

\begin{figure}
\begin{center}
\includegraphics[width=0.49\textwidth]{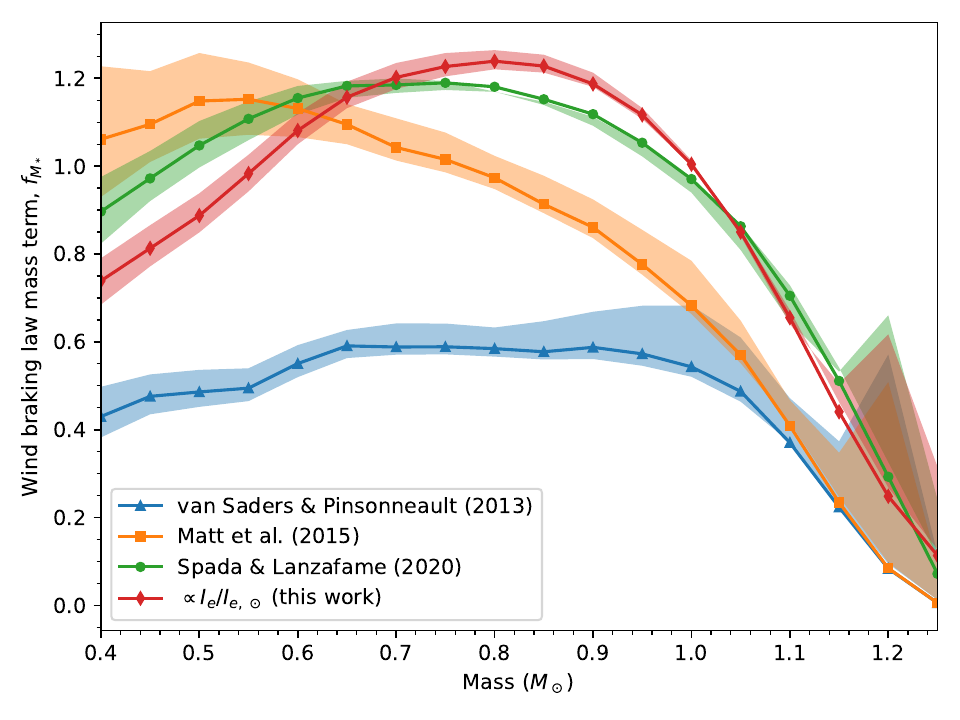}
\caption{Prescriptions for the mass dependence of the wind braking law, $f_{M_*}$, considered in our models. The range of variability due to stellar evolution in the age range spanned by our models ($\approx 0.1$--$4$ Gyr) is also shown. The increase in the variability width at the two ends of the mass range is due to the prolonged pre-main sequence phase, and to the earlier onset of the post-main sequence phase, respectively, in comparison with the stars in the middle of the mass range.}
\label{fig:mass_dep}
\end{center}
\end{figure}

Several $f_{M_*}$ prescriptions exist in the literature, whose motivation range from purely theoretical models, to fits of theoretical models, to semi-empirical, to purely empirical considerations. 
In this work, we discuss models implementing the following mass dependence terms of the wind braking law:
\begin{subequations} 
\label{eq:mass_dep}
{\small
\begin{align}
f_{M_*} = & \left(\frac{R_*}{R_\odot}\right)^{3.1} \left(\frac{M_*}{M_\odot}\right)^{-0.22} \left(\frac{L_*}{L_\odot}\right)^{0.56} \left(\frac{P_s}{P_{s, \odot}}\right)^{0.44} \left(\frac{\tau_c}{\tau_{c, \odot}}\right)^2;
\label{eq:mass_dep_vs13} 
\\
f_{M_*} = & \left(\frac{R_*}{R_\odot}\right)^{3.1} \left(\frac{M_*}{M_\odot}\right)^{0.5} \left(\frac{\tau_c}{\tau_{c, \odot}}\right)^2;
\label{eq:mass_dep_ma15} 
\\
f_{M_*} = & \left(\frac{I_*}{I_\odot}\right) \left(\frac{\tau_c}{\tau_{c, \odot}}\right);
\label{eq:mass_dep_sl20} 
\\
f_{M_*} = & \left(\frac{I_e}{I_{\rm e, \odot}}\right).
\label{eq:mass_dep_ienv} 
\end{align}
}
\end{subequations}
In the equations above, $L_*$ and $P_s$ are the luminosity and the photospheric pressure of the star, respectively; $\tau_c$ is the {global} convective turnover timescale \footnote{The ``global'' (or ``nonlocal'') convective turnover timescale is calculated as the integral of the reciprocal of the local convective velocity, $v_c$, extended from the bottom of the convection zone to the surface: $\tau_c = \int_{R_c}^{R_*} \frac{dr}{v_c}$ \citep{Kim_Demarque:1996}.}; $I_* = I_e + I_c$ is the moment of inertia of the whole star.
Each of these quantities enters $f_{M_*}$ scaled over its respective solar counterpart, which is indicated with the additional subscript ``$\odot$''.
During the integration of equations \eqref{tzm}, the dependence on these additional structure quantities is handled in the same way as discussed above for the rotation-related parameters.

Detailed discussions of equations \eqref{eq:mass_dep_vs13}--\eqref{eq:mass_dep_sl20} can be found in \citet{Matt_ea:2012} and \citet{vanSaders_Pinsonneault:2013}; \citet{Matt_ea:2015}; and \citetalias{SL20}, respectively. 
Equation \eqref{eq:mass_dep_ienv} is original to the present work, and will be further discussed below.

The functional dependence of $f_{M_*}$ on the mass of the star according to equations \eqref{eq:mass_dep} is illustrated in Figure~\ref{fig:mass_dep}, which also shows the range of variation due to stellar evolution between $0.1$ and $4.0$ Gyr.
It is clear from the Figure that the indirect time dependence introduced by stellar evolution is quite modest, except at the two extremes of the mass range.
The widening of the $f_{M_*}$ time variation at $M_* \lesssim 0.5\, M_\odot$ and at $M_* \gtrsim 1.15\, M_\odot$ is due to the non-negligible overlap of the pre-main sequence and of the post-main sequence phases, respectively, with the age range considered here.

The following quantities are treated as adjustable parameters in our rotational evolution model:
\begin{itemize} 
\item The initial rotation period $P_0$;
\item The circumstellar disk lifetime $\tau_{\rm disk}$; 
\item The scaling constant $K_w$ in the wind braking law; 
\item The rotational coupling timescale $\tau_{\rm cpl}$.
\end{itemize}

The effect of the parameters $P_0$ and $\tau_{\rm disk}$ is mostly limited to the very early stages of the rotational evolution, which should be practically forgotten as soon as stars settle on the \srs{}. 
We thus treat $P_0$ and $\tau_{\rm disk}$ as nuisance parameters, which we fit independently of stellar mass.
In contrast, based on the results of \citetalias{LS15, SL20}, we anticipate that significantly mass-dependent  $K_w$ and $\tau_{\rm cpl}$ will be required to reproduce the observations, for different reasons.
The dependence of $K_w$ on stellar mass can be considered as an empirical correction to the shortcomings of the assumed $f_{M_*}$ prescription.
A mass-dependent $\tau_{\rm cpl}$, on the other hand, is critical to properly model the stalled magnetic braking phase.

\subsection{Fitting the parameters of the two-zone model}
\label{tzm_fit}

Once the stellar evolution tracks and the input parameters are specified, our rotational evolution model of the \srs{} is completely determined.
We can thus construct snapshots of the sequence at the nominal ages of the open clusters in our catalog (see Table~\ref{table_oc}).
In practice, we do so by calculating the rotational evolution of stars in the mass range $0.4$--$1.25 \, M_\odot$, in steps of $0.05\, M_\odot$, independently for each mass; as a result, our calculated \srs{} is sampled at $18$ points equally spaced in mass.

Each of the modeled sequence snapshots is compared with its empirical counterpart of the same age, obtained from the non-parametric fits described in Section~\ref{nonparmfit}. 
In this way, we form the residuals, in the sense of observed-minus-calculated (O-C) differences, normalized to the standard deviations of the non-parametric fits.
Our goodness-of-fit metric is defined as a standard chi-square in terms of these residuals:
\begin{equation}
\label{chisq}
\chi^2 = \sum_{i=1}^N \frac{(O_i-C_i)^2}{\sigma_i^2},
\end{equation}
where the index $i$ runs over the sampled masses in the sequence, and in turn, over the ages of the clusters in our catalog.
It should be noted that the empirical \srs{} does not exist in the full mass range at all ages, due both  to incomplete data coverage and to the mass-dependence of the timescale of convergence on the sequence (i.e., lower mass stars converge to the sequence at later ages than their more massive counterparts).
As a result, the total number of empirical constraints on the fit is $N=114$, out of the maximum possible $144$ values ($=18$ masses $\times$ $8$ clusters).

We determine the best-fitting values of the parameters using a standard least-squares approach, as implemented in the \texttt{least\_squares} function of the \texttt{scipy.optimize} package.
As explained above, while the parameters $P_0$ and $\tau_{\rm disk}$ are assumed to be independent of the mass, the parameters $K_w$ and $\tau_{\rm cpl}$ are allowed to take different values for each stellar mass.
The total number of free parameters in our fits is therefore $p = 2 + 2 \times 18 = 38$, independent of the prescription chosen for $f(M_*)$.

\begin{figure*}
\begin{center}
\includegraphics[width=0.95\textwidth]{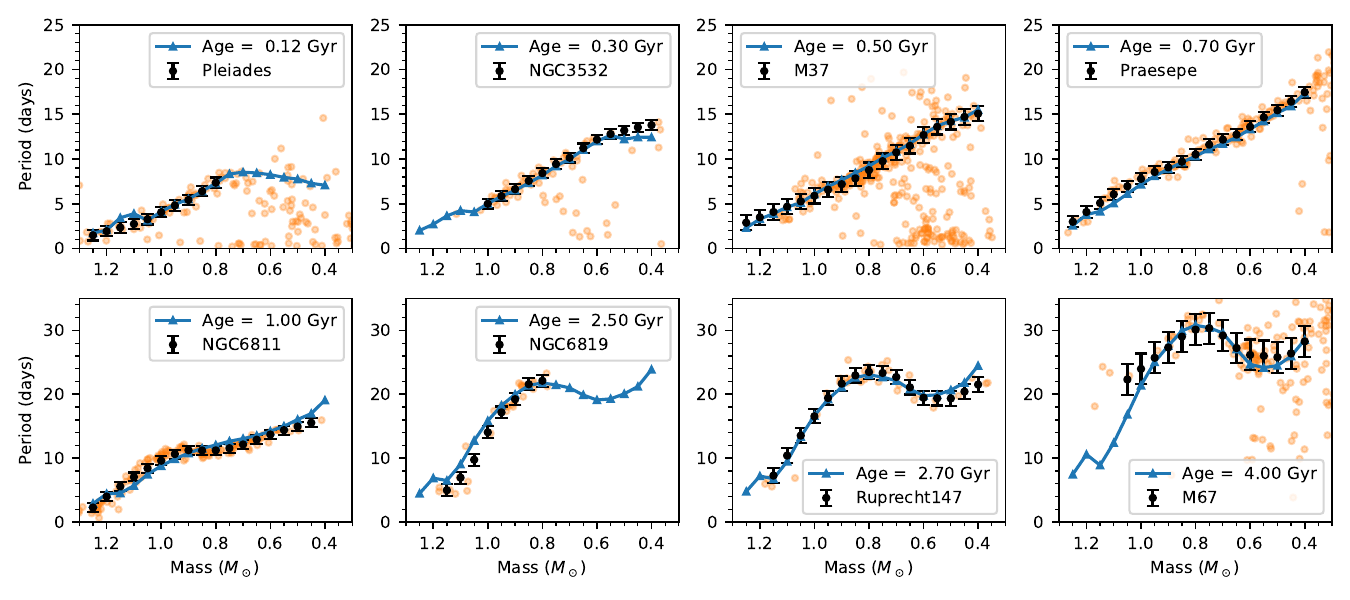}
\caption{Rotational isochrones from our model ``ienv'' (blue lines with triangles) compared with the non-parametric fit of the \srs{} (black circles with error bars) in the mass--period diagram of the clusters in our compilation. Individual stars are also shown as orange circles.}
\label{fig:cpd}
\end{center}
\end{figure*}

\begin{figure*}
\begin{center}
\includegraphics[width=\textwidth]{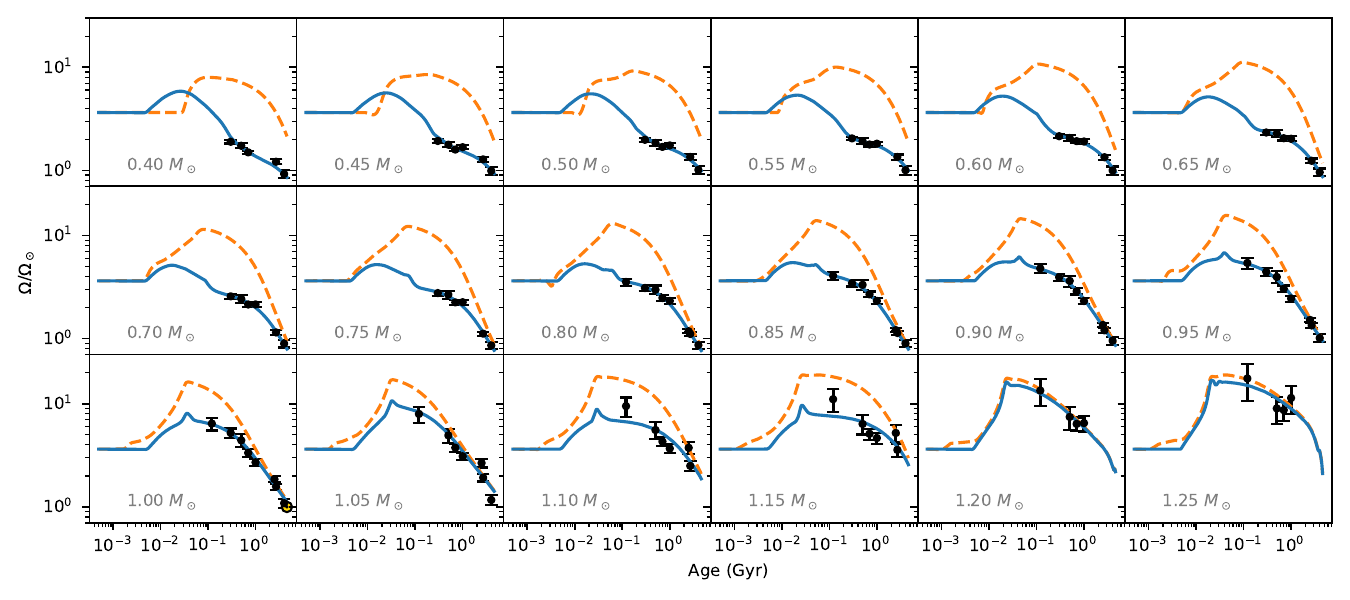}
\caption{Rotational evolution according to model ``ienv'' for stars of mass $0.4$--$1.25\, M_\odot$. In each panel, the angular velocity of the convective envelope is plotted as a blue solid line, the angular velocity of the radiative interior is plotted as a dashed orange line, and the constraints from the non-parametric fits of the \srs{} are shown as black circles with error bars (the surface period of the Sun is also shown in the appropriate panel).}
\label{fig:evl}
\end{center}
\end{figure*}

\section{Results}
\label{results}

We fitted our rotational evolution models to the empirical constraints on the \srs{} discussed in Section~\ref{nonparmfit}, using the least-squares procedure described in Section~\ref{tzm_fit}.
The results of these fits are summarized in Table~\ref{tab:fit_results}.
The first four entries in the Table, models ``vs13'', ``ma15'', ``sl20'', and ``ienv'', differ in the mass dependence of the wind braking law, $f_{M_*}$, while model ``bpl'' is qualitatively different, and will be  discussed in detail in Section~\ref{wind_mass_dep}.
The reduced chi-square is $\chi_\nu^2 = \chi^2/(N-p)$, where $N$ and $p$ are the number of constraints and free parameters in the fit for each model, respectively.

Besides the chi-square and reduced chi-square values, Table~\ref{tab:fit_results} lists the best-fitting $P_0$ and $\tau_{\rm disk}$ for each model. 
The mass-dependent best-fitting $K_w$ and $\tau_{\rm cpl}$ are discussed in detail in the next two sections. 
To illustrate our results, we plot the rotational isochrones for model ``ienv'' ($f_{M_*} = {I_e}/{I_{e, \odot}}$), together with the individual data and the non-parametric fit of the \srs{} in Figure~\ref{fig:cpd}. 
The rotational evolution of individual stars in our mass range according to the same model are plotted in Figure~\ref{fig:evl}.

Comparing the chi-square values in Table~\ref{tab:fit_results}, we see that the most satisfactory fit is obtained with $f_{M_*} = {I_e}/{I_{e, \odot}}$, equation \eqref{eq:mass_dep_ienv}. 
We propose this novel form of $f_{M_*}$ on mostly empirical grounds. 
We note (see Figure~\ref{fig:mass_dep}) the overall similarity of the mass dependence introduced by equations \eqref{eq:mass_dep_ienv} and \eqref{eq:mass_dep_sl20}, the latter being a modification suggested by \citetalias{LS15} of an earlier form proposed in turn by \citet{Barnes_Kim:2010}.
An independent argument in support of equation \eqref{eq:mass_dep_ienv} is offered in Section~\ref{wind_mass_dep}. 

\begin{table}[h]
\caption{Summary of the best-fitting parameters in our rotational evolution models (see text for details).}
\begin{center}
\begin{tabular}{ccccccccc}
\hline
\hline
Model & $P_0$  & $\tau_{\rm disk}$ & $f_{M_*}$ & $\chi^2$ & $\chi_\nu^2$ \\
 & (days) & (Myr) & & &  \\
\hline
vs13 & $7.04$ & $4.54$ & eq. \eqref{eq:mass_dep_vs13}    & $152.6$ &  $2.01$ \\
ma15 & $4.01$ & $3.07$ & eq. \eqref{eq:mass_dep_ma15} & $141.3$ &  $1.86$ \\
sl20 & $7.72$ & $4.82$ & eq. \eqref{eq:mass_dep_sl20}     & $120.7$ &  $1.59$ \\
ienv & $7.18$ & $4.80$ & eq. \eqref{eq:mass_dep_ienv}     & $108.2$ &  $1.42$ \\
\hline
bpl  & $10.1$ & $4.57$ & 1.0 & $ 97.7$  & $1.04$ \\
\hline
\end{tabular}
\end{center}
\label{tab:fit_results}
\end{table}

\begin{figure}
\begin{center}
\includegraphics[width=0.49\textwidth]{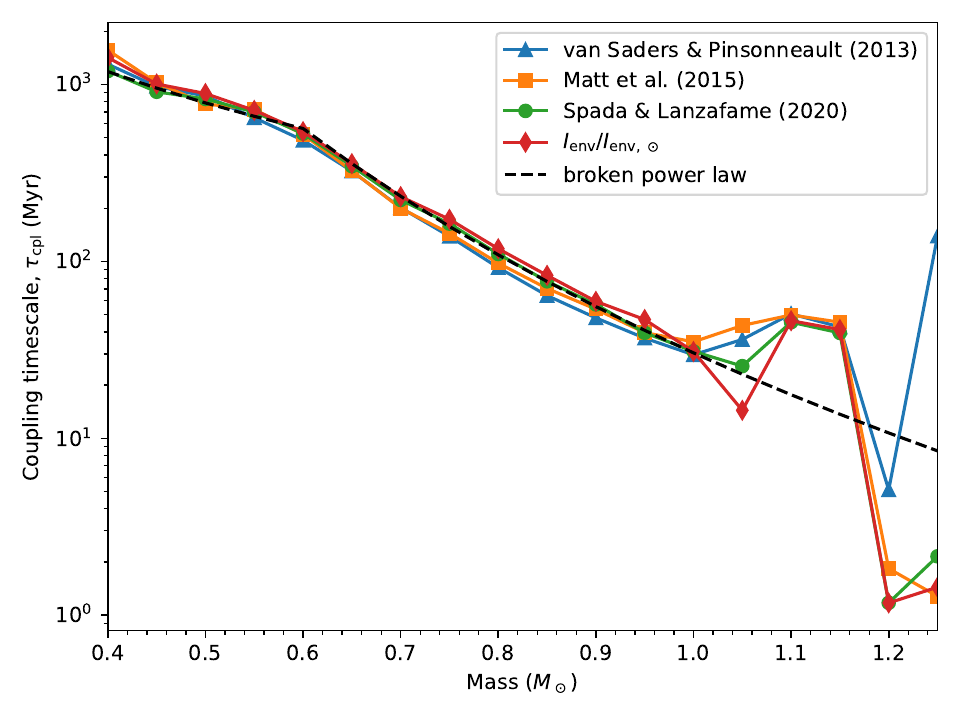}
\caption{Best-fitting $\tau_{\rm cpl}$ resulting from the least-squares optimization procedure described in Section~\ref{tzm_fit}. The dashed black line represents a best-fit of the common $\tau_{\rm cpl}$ behavior with a broken power law (see equation \ref{eq:bpl} and the discussion in the text).}
\label{fig:tauc}
\end{center}
\end{figure}

\subsection{The rotational coupling timescale}
\label{rotcoupl}

The best-fitting values of the rotational coupling timescale $\tau_{\rm cpl}$ turn out to be fairly insensitive to the choice of $f_{M_*}$ (see Figure~\ref{fig:tauc}).
This is in itself an important result, because a strong correlation between the wind braking term and the angular momentum redistribution term in equations \eqref{tzm} could not have been ruled out a priori. 

\begin{figure}
\begin{center}
\includegraphics[width=0.49\textwidth]{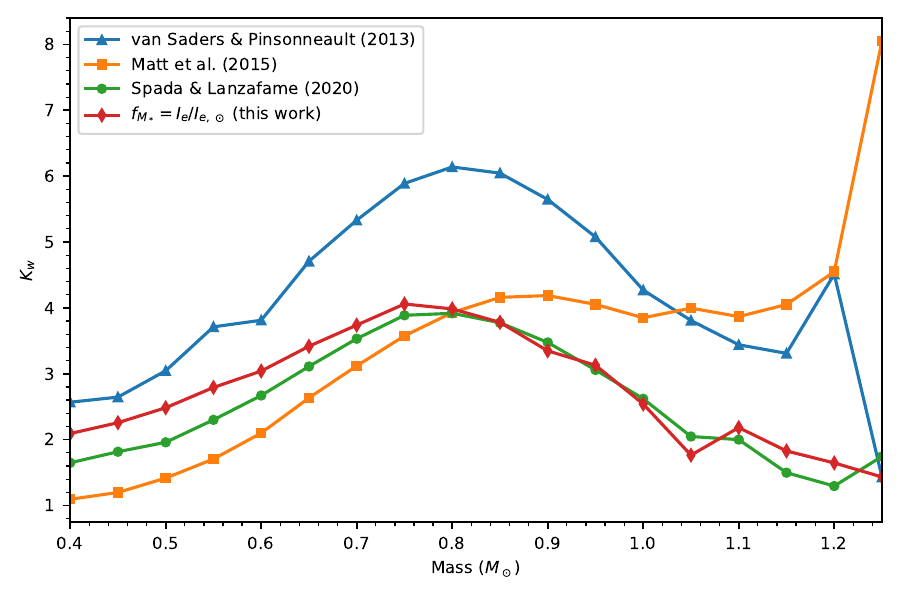}
\caption{Best-fitting $K_w$ resulting from the least-squares optimization procedure described in Section~\ref{tzm_fit}. }
\label{fig:Kw}
\end{center}
\end{figure}

Notably, the largest differences among the $\tau_{\rm cpl}$ fitted with different $f_{M_*}$ terms are confined to $M_* \gtrsim 1.1\, M_\odot$.
In this mass subrange, the best-fitting $\tau_{\rm cpl}$ exhibits relatively large, non-smooth fluctuations.
A similar situation, albeit characterized by more modest amplitudes, also arises for $M_* \approx 0.4\, M_\odot$.
In general, we can expect our fits to be the least robust at the two ends of the mass range, since in those regimes i) the empirical constraints on the \srs{} are often missing in the color--period diagrams of the clusters in our catalog; ii) the largest variations of the stellar parameters due to stellar evolution occur.
Indeed, we observe a similar effect in the $K_w$ best-fitting values (see Figure~\ref{fig:Kw}).

{
When the coupling timescale is sufficiently short, the star rotates essentially as a solid body throughout most of its evolution. In this regime, the rotational evolution model becomes increasingly insensitive to the precise value of $\tau_{\rm cpl}$.
Best-fitting values of $\tau_{\rm cpl} \lesssim 10$ Myr are therefore intrinsically poorly constrained, which explains, at least in part, the large relative variations obtained for $M_* \gtrsim 1.1\, M_\odot$. 
}

For these reasons, it is meaningful and practically advantageous to extract the general behavior of $\tau_{\rm cpl}$, while filtering out the noisy fluctuations at the high end of the mass range.
For this purpose we represent the mass dependence of the rotational coupling timescale obtained from our fits with a broken power law of the form:
\begin{equation}
\label{eq:bpl}
\tau_{\rm cpl} = \tau_0
\left\{
\begin{array}{cc}
\left(\frac{M_*}{M_b}\right)^{-\alpha_1}, & M_* < M_b \\
\left(\frac{M_*}{M_b}\right)^{-\alpha_2}, & M_* > M_b \\
\end{array}
\right.
\end{equation}
By visual inspection, we set the mass at which the exponent break occurs to $M_b = 0.60\, M_\odot$, and we derive the parameters $\tau_0$, $\alpha_1$, and $\alpha_2$ from a standard curve fit performed with \texttt{scipy.optimize.curve\_fit}.
We obtain:
\begin{equation}
\label{eq:bpl_parms}
\tau_0 = 565. \pm 15. \ {\rm Myr}; \ \ \alpha_1 = 1.83 \pm 0.09; \ \ \alpha_2 = 5.72 \pm 0.32
\end{equation}
The resulting curve, plotted in Figure~\ref{fig:tauc} as a dashed black line, provides a very good parametric representation of our empirically derived mass dependence of the rotational coupling timescale.
For reference, this parametrization corresponds to $\tau_{\rm cpl}=30.4 \pm 5.0$ Myr for a $1.0 \, M_\odot$ star.
Along with the higher mass regime exponent $\alpha_2$, this value can be directly compared with the best-fitting values in the power law function proposed by \citetalias{SL20}: $\tau_{\rm cpl, \odot}^{\rm SL20} = 22$ Myr, $\alpha^{\rm SL20} = 5.6$.

Our current broken power law model of $\tau_{\rm cpl}$ thus extends our previous results, while being completely consistent with it within their common range of applicability (although the \citetalias{SL20} power law parameters were presented without formal error bars, it is reasonable to assume that their uncertainties should be comparable to those of our current broken power law fit).
The agreement of our present and past results is encouraging, also considering that the power-law behavior of $\tau_{\rm cpl}$ found by \citetalias{LS15} had been confirmed, both qualitatively and quantitatively, by a completely independent study \citep{Somers_Pinsonneault:2016}.

Using the broken power law parametrization for the rotational coupling timescale {not only filters out the large relative flucturations of $\tau_{\rm cpl}$ at the high mass end of the range, but} can also reduce the total number of free parameters in the least-squares fit: the corresponding model is referred to as ``bpl'' in Table~\ref{tab:fit_results} and in the following (see the next Section for details).

\subsection{The wind braking law mass dependence}
\label{wind_mass_dep}

Figure~\ref{fig:Kw} compares the best-fitting $K_w$'s obtained for each stellar mass with different choices of the mass dependence term $f_{M_*}$, as reported in Table~\ref{tab:fit_results}.
Ideally, the $f_{M_*}$ term should fully account for the mass dependence of the wind braking law, resulting in optimized $K_w$ nearly independent of mass. 
In other words, the smallness of the mass dependence found in the fitted $K_w$ can be taken as a measure of the quality of the $f_{M_*}$ prescription used (cf. the discussion in \citetalias{LS15}).
We see from Figure~\ref{fig:Kw} that models ``sl20'' and ``ienv'' are the best performing with respect to this criterion.

To take advantage of the regularized empirical mass dependence of $\tau_{\rm cpl}$ as a broken power law expressed by equation \eqref{eq:bpl} (see Section \ref{rotcoupl}), we have performed an additional fit, which we refer to as model ``bpl'' in the following.
This fit is qualitatively different from the others in two respects:
\begin{enumerate} 
\item 
The rotational coupling timescale is not freely adjusted, but prescribed according to equations \eqref{eq:bpl} with the coefficients \eqref{eq:bpl_parms}; 
\item
We set $f_{M_*}=1$, so that the mass dependence of the wind braking law is fully encoded in the adjusted $K_w$'s.
\end{enumerate}

\begin{figure}
\begin{center}
\includegraphics[width=0.49\textwidth]{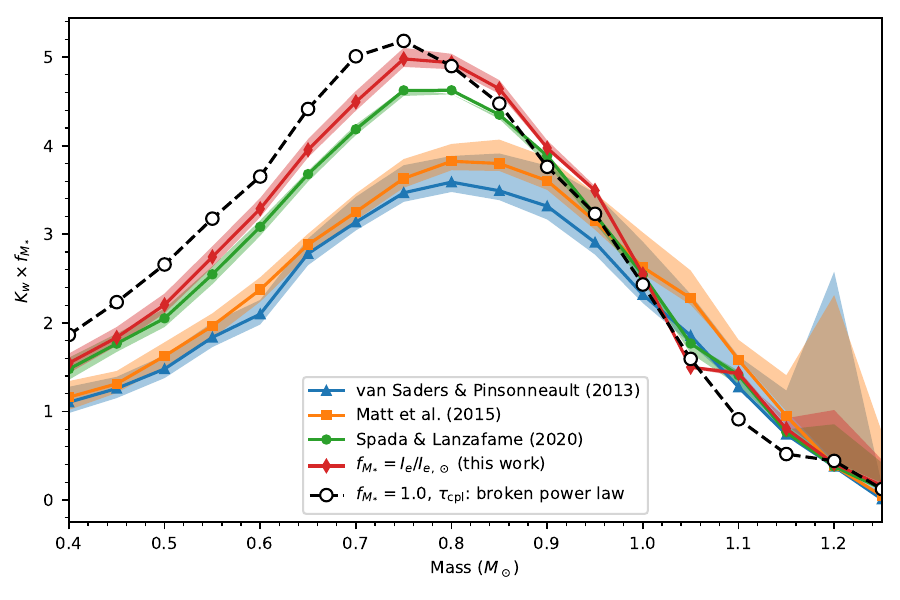}
\caption{Similar to Figure~\ref{fig:Kw}, but showing the product $K_w \times f_{M_*}$, and including the fit obtained with model ``bpl''. Note the very close agreement between the latter and model ``ienv''.}
\label{fig:Kw_times_f}
\end{center}
\end{figure}

Figure~\ref{fig:Kw_times_f} shows the fit corresponding to model ``bpl'' along with the ones based on the $f_{M_*}$ prescriptions in equations \eqref{eq:mass_dep} (dashed black line with open circles; cf. the corresponding entry in Table~\ref{tab:fit_results}).
In order to make the comparison in Figure~\ref{fig:Kw_times_f} meaningful, in this case the product $K_w \times f_{M_*}$ is plotted for each model.

Because $\tau_{\rm cpl}$ is held fixed, model ``bpl'' contains $p=2+18=20$ adjustable parameters (i.e., $P_0$, $\tau_{\rm disk}$, and the $18$ mass-dependent $K_w$ values).
This is therefore the most economical of our models, and thus the closest to extracting an empirical $K_w \times f_{M_*}$ from our data sample.
Of course, even in this case, the fitted mass dependence cannot be considered, {\it sensu stricto}, fully empirical, since the fitting procedure  introduces some model-dependent assumptions. 
Another advantage of the approach in model ``bpl'' is that the fluctuations of $\tau_{\rm cpl}$ at the two extremes of the mass range are effectively removed.

The ``semi-empirical'' fit of $K_w \times f_{M_*}$ obtained with model ``bpl'' is therefore a valuable reference for comparison, and a useful guide to formulate a wind braking law with an improved mass dependence term.
In particular, since, as stated above, the ideal mass term $\tilde f_{M_*}$ should be the one for which the corresponding best-fitting $\tilde K_w$ is independent of the stellar mass, we can speculate that:
\begin{equation*}
\left.
\begin{array}{cc}
(\tilde K_w \times \tilde f_{M_*}) = (K_w^{\rm bpl} \times 1);  
\\ 
\\
\tilde K_w \approx {\rm const.} 
\end{array}
\right\}
\Rightarrow 
\tilde f_{M_*} \propto K_w^{\rm bpl},
\end{equation*}
where $K_w^{\rm bpl}(M_*)$ are the best-fitting values of the wind braking law constant from the fit of model ``bpl''.

In other words, the quantites $K_w^{\rm bpl}(M_*)$ can be used to constrain the optimal mass dependence of $f_{M_*}$ within an overall multiplicative constant.
As a step in this direction, we note that our model ``ienv'' results in a $K_w \times f_{M_*}$ which traces most closely the semi-empirical one, the difference between the two being at most $20\%$ between $0.4$ and $1.1\, M_\odot$.

\begin{figure*}
\begin{center}
\includegraphics[width=\textwidth]{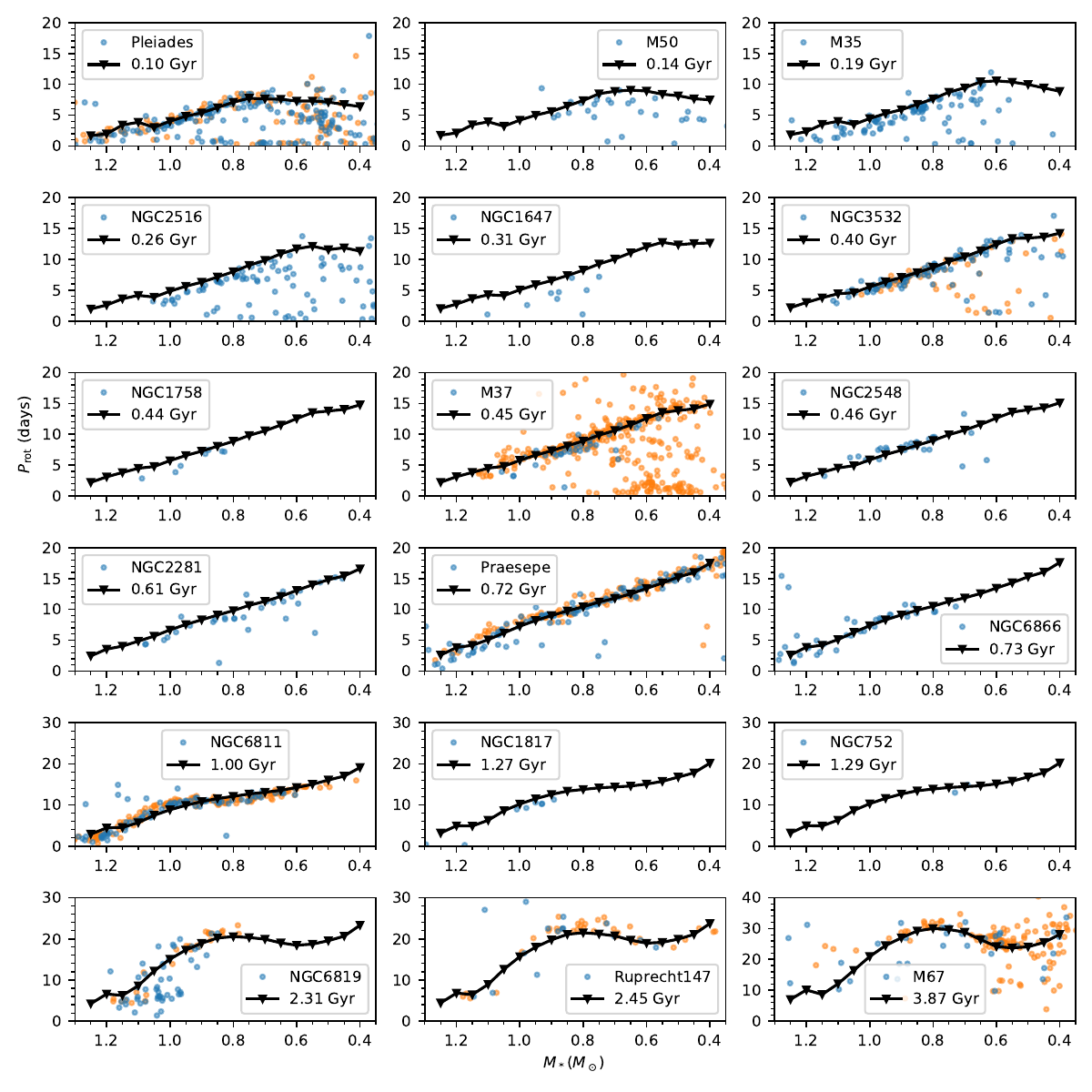}
\caption{Comparison between the \texttt{ChronoFlow} catalog (\citealt{Van-Lane_ea:2025}, blue), our data compilation (orange), and our best model, using $f_{M_*}$ in equation \eqref{eq:mass_dep_ienv} (black triangles with solid lines). Note that the rotational isochrones shown are calculated at the fiducial ages of the \texttt{ChronoFlow} catalog, which are different from ours for some clusters (see text for details).}
\label{fig:chronoflow}
\end{center}
\end{figure*}

\subsection{A posteriori validation} 
\label{chronoflow} 
 
The catalog of rotation period measurement recently assembled by \citet{Van-Lane_ea:2025} in support of their gyrochronology model \texttt{ChronoFlow} provides the opportunity for an independent validation of our work.
Indeed, our literature compilation, including the selection of open clusters, choice of fiducial ages, sources of rotation periods, Gaia colors, cluster membership information, detailed in Section~\ref{data}, is completely independent of theirs.
In addition, since the \texttt{ChronoFlow} catalog features a more extensive compilation of open clusters than ours, we can use it to test our rotational models on data which were not included in the fitting procedure.
 
The comparison is summarized in the plots of Figure~\ref{fig:chronoflow}.
In this Figure, the data from the \texttt{ChronoFlow} catalog is plotted in blue, while the data from our catalog, where available, is shown in orange; rotational isochrones from our model ``ienv'' are plotted as solid black lines with triangles.
For internal consistency of the comparison, the isochrones shown in each panel of Figure~\ref{fig:chronoflow} are calculated at the fiducial ages of the clusters adopted in the \texttt{ChronoFlow} catalog, which differ from our nominal ages (Table~\ref{table_oc}) as follows: Pleiades: $100$ vs $120$ Myr; NGC 3532: $400$ vs $300$ Myr; M37: $450$ vs $500$ Myr; Praesepe: $720$ vs $700$ Myr; NGC 6819: $2.31$ vs $2.5$ Gyr; Ruprecht 147: $2.45$ vs $2.7$ Gyr; M67: $3.87$ vs $4.0$ Gyr.
This comparison thus seems to indicate that moderate changes in the nominal ages do not significantly affect the final results.

The agreement of the model with the data, in particular with the \srs{} or at least the upper envelope of the rotation periods distribution of each cluster, is very good.
This comparison strengthens our confidence in our modeling approach, and in particular in our new proposal for the wind mass dependence prescription, $f_{M_*} \propto I_e$.

\section{Discussion}
\label{discussion}

In this paper, we adopted a modeling framework for the rotational evolution of solar-like stars which is capable of reproducing the evolution of the \srs{} as observed in Galactic open clusters.
With the present revision of the modeling parameters, our models are applicable in the entire mass range encompassing stars with solar-like interior structure (i.e., a radiative interior and a convective envelope) during their main sequence phase, namely, $0.4$--$1.25\, M_\odot$. 
Our models reproduce satisfactorily the evolution of the \srs{} in the age interval from $\approx 100$ Myr to $4$ Gyr.

A key feature of our models is that they account for the differential rotation between the radiative interior and the convective envelope, which develops as the angular momentum redistribution from the interior lags behind the loss at the surface via magnetic braking.
The temporary rotational decoupling, and subsequent re-coupling, is indeed considered the leading explanation for the observed non-monotonic rotational evolution, or stalled braking, first observed between the Praesepe and NGC6811 clusters, i.e., between $0.7$ and $1.0$ Gyr, for stars of approximately $0.8\, M_\odot$ \citep{Curtis_ea:2019, Curtis_ea:2020, Angus_ea:2020, Gordon_ea:2021, Santos_ea:2025}.

The rotational coupling timescale $\tau_{\rm cpl}$, which parametrizes the angular momentum transport efficiency in our models, was found by \citetalias{LS15} and \citetalias{SL20} to be strongly mass-dependent, following a power law with negative exponent.
This result is confirmed by our present fit, and the revised coefficients are in good agreement with those found in our previous work.

It was further pointed out by \citetalias{SL20} that, as a consequence of the mass dependence of $\tau_{\rm cpl}$, stars of progressively lower mass should experience the stalling of their rotational braking at increasingly later ages, since the longer $\tau_{\rm cpl}$ (see Figure~\ref{fig:tauc}) results in progressively later recoupling.
This prediction appears to be confirmed by the most recent observations. 
In clusters of age $\gtrsim 2$ Gyr, the empirical \srs{} features a characteristic dip at masses below $0.8\, M_\odot$. 
The minimum of this dip shifts to a lower mass as the age of the cluster increases, as can be seen from the comparison of NGC 6811, Ruprecht 147, and M67 (cf. Figure~\ref{fig:cpd}).
This feature is recovered by our models, and we speculate that it is the signature of stalled braking occurring at increasingly lower masses as a consequence of the mass-dependence of $\tau_{\rm cpl}$. 
This ``stalling wave'' was predicted by \citetalias{SL20}, and is reproduced by our current models.

Our results provide semi-empirical constraints on the mass dependence of the two key processes driving the rotational evolution of solar-like stars: the angular momentum transport in the interior and the wind braking at the surface (see Sections \ref{rotcoupl} and \ref{wind_mass_dep}). 
While this scaling may be of interest in exploring candidate mechanisms for angular momentum transport in stellar radiative interiors \citep[e.g.,][]{Fuller_ea:2014, Eggenberger_ea:2022, Meduri_ea:2024}, we emphasize that our results do not identify the underlying physical process, and should not be interpreted as providing a complete theoretical explanation. 
Similarly, the empirical $K_w \times f_{M_*}$ relation can serve as a practical guide for calibrating MHD wind models as a function of mass.

From a practical perspective, our models represent a step toward a physics-informed gyrochronology relation, complementing purely empirical or data-driven approaches \citep[e.g.,][]{Barnes:2010, Van-Lane_ea:2025}.

Because our models adopt the two-zone framework, their extension to stars without a solar-like internal structure (most notably, fully convective stars, $M_* \lesssim 0.35\, M_\odot$) is not straightforward. 
In this regime the physical assumptions underlying the core-envelope decoupling formalism are no longer valid.
In this context, recent work by \citet{Chiti_ea:2024} indicates that rotation periods increase sharply across a narrow effective temperature interval near the fully convective boundary. 
This behavior may reflect the structural changes (e.g., convection zone depth, convective turnover timescale) associated with the transition, and its dependence on mass, metallicity, and age, potentially linked to the $^3$He-burning instability discovered by \citet{vanSaders_Pinsonneault:2012}. 
Clearly, models based on different assumptions and likely belonging to a different model class are required to incorporate the fully convective regime within a unified framework for solar-like and low-mass stellar rotational evolution \citep[see also][]{Lu_ea:2024}.

The present study focuses on the Skumanich-like regime of rotational evolution, in which the spin down by magnetic wind braking continues indefinitely. 
This framework has proven robust up to about solar age, and it remains the natural reference against which new developments are measured. 
Currently, the $4$ Gyr-old open cluster M67 provides the oldest available rotational constraints, and its stellar population is consistent with traditional spin-down laws at least up to that age.

Recent observations of field stars suggest a departure from standard magnetic braking behavior, with evidence for an abrupt reduction of the wind torque as the Rossby number (the ratio between the rotation period and the convective turnover timescale) approaches a critical value \citep{vanSaders_ea:2016, Metcalfe_ea:2025}. 
Because the relevant control parameter is the Rossby number, the age at which weakened magnetic braking sets in is intrinsically mass-dependent. 
It cannot therefore be ruled out that some of the most massive stars in the open clusters considered here may already lie in this regime. 
In this work, however, we do not attempt a quantitative assessment, in part owing to the multiple, non-equivalent definitions of the Rossby number used across theoretical and observational studies.

Although our models do not attempt to capture {the weakened magnetic braking} regime, they provide a controlled baseline that can be extended once the relevant physical processes (such as changes in coronal magnetic topology, dynamo action, or the coupling between the stellar interior and wind) are better understood. 
Future work may therefore build upon the present results by introducing prescriptions for weakened braking at critical Rossby number, guided by asteroseismic and Gaia-based constraints. 
In this way, the applicability of our approach could be extended from the well-calibrated young and middle-aged populations to encompass the full main-sequence lifetime of solar-like stars, and possibly beyond.

\section{Conclusions}
\label{conclusions}

The present paper extends and complements the results of our previous work, by bringing to fruition the latest observational constraints recently obtained through ground- and space-based photometric surveys of stellar rotation.

Our improved models can satisfactorily reproduce the observed evolution of the \srs{} in open clusters from $100$ Myr to $4$ Gyr for the entire range of solar-like stars ($0.4$--$1.25\, M_\odot$).
{Our physically motivated parametric model} can constrain the dependence on stellar mass of the two key processes which drive the rotational evolution of solar-like stars, namely, the wind braking at the surface and the angular momentum transport in the interior.

We find that the rotational coupling timescale as a function of stellar mass follows very closely a broken power law relation, with the exponent change occurring at $0.6 \, M_\odot$, i.e., at the transition from late to early K-type stars. 
{
Importantly, the rotational coupling timescale as a function of stellar mass follows this broken power-law relation independently of the choice of the wind braking law, demonstrating that the coupling mass dependence is robust to variations in surface torque prescriptions.
}
The optimized parameters of the broken power law are in very good agreement with our previous findings, which had also been independently confirmed by other authors.

By setting the rotational coupling timescale according to the optimized broken power law, we can reduce the number of adjustable parameters in the rotational evolution model, and derive a ``semi-empirical'' mass dependence of the wind braking law, i.e., one obtained with a minimum of model-dependent assumptions.

Finally, we propose a novel formulation for the mass dependence of the wind braking law, proportional to the moment of inertia of the convective envelope of the star.
Such formulation results in the best agreement with the observations among the functional wind braking laws we tested, and is indirectly supported by its close resemblance with the semi-empirical mass scaling.

\begin{acknowledgements}
This research has made use of the following software: NumPy \citep{Harris_ea:2020}, SciPy \citep{Virtanen_ea:2020}, Matplotlib \citep{Hunter_ea:2007}. FS is funded by the European Union – NextGenerationEU RRF M4C2 1.1 No.2022HY2NSX, “CHRONOS: adjusting the clock(s) to unveil the CHRONO-chemo-dynamical Structure of the Galaxy” (PI: S. Cassisi). \\
FS gratefully acknowledges the inspiring mentorship and personal kindness of the late Pierre Demarque, and the friendship and professional support of the late Patrick Gaulme, both extended over many years of close scientific collaboration.
\end{acknowledgements}

\bibliographystyle{aa}

\clearpage
\appendix

\section{Rotational isochrones}

 \begin{figure}
\begin{center}
\includegraphics[width=0.49\textwidth]{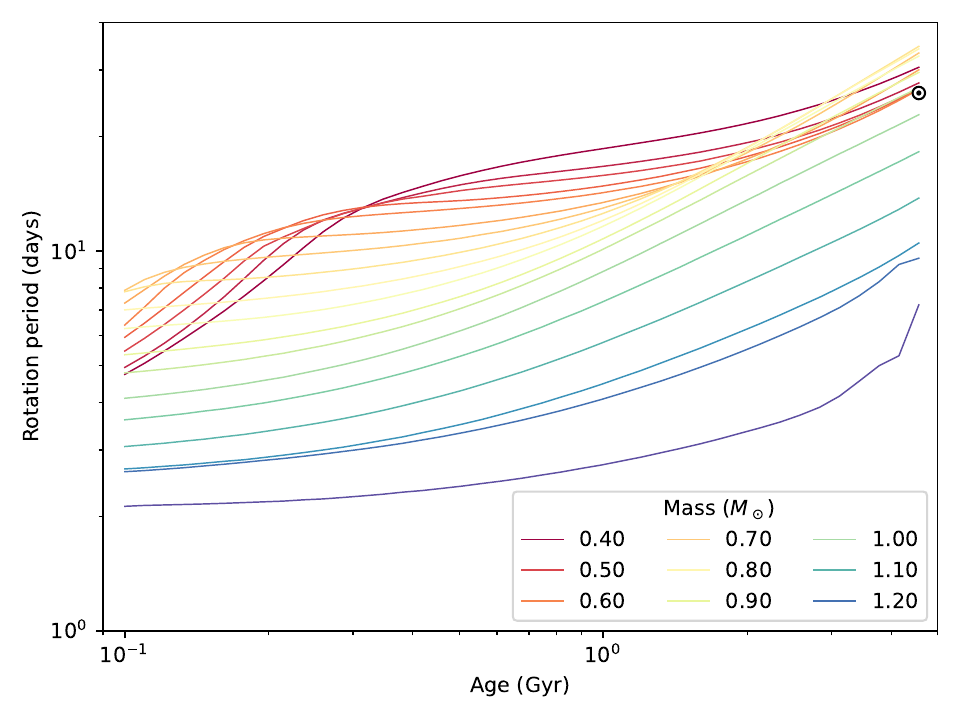}
\caption{Evolution of the \srs{} according to model ``bpl'': rotational history for models of selected mass.}
\label{fig:piso_sbpl1}
\end{center}
\end{figure}

The Figures and the Table in this Appendix are meant to provide a qualitative and quantitative impression of the rotational evolution of the \srs{} according to model ``bpl''.
The main features of the rotational isochrones discussed in this Section are shared by those obtained with model ``ienv''.
For a detailed description of these models, see Section~\ref{results}.
 
Figure~\ref{fig:piso_sbpl1} shows the evolution of the surface rotation period as a function of the stellar mass, which is the direct output of our model; the Sun is also plotted with its usual symbol. 
It should be noted that our models are not calibrated to reproduce the solar rotation period exactly.
The model at $0.95 \, M_\odot$ turns out to be the one closest to the solar period ($26.09$ days) at $4.57$ Gyr.
This is not surprising, since the solar rotation period was not included as a constraint in our fit. 
 
\begin{figure}[h!]
\begin{center}
\includegraphics[width=0.49\textwidth]{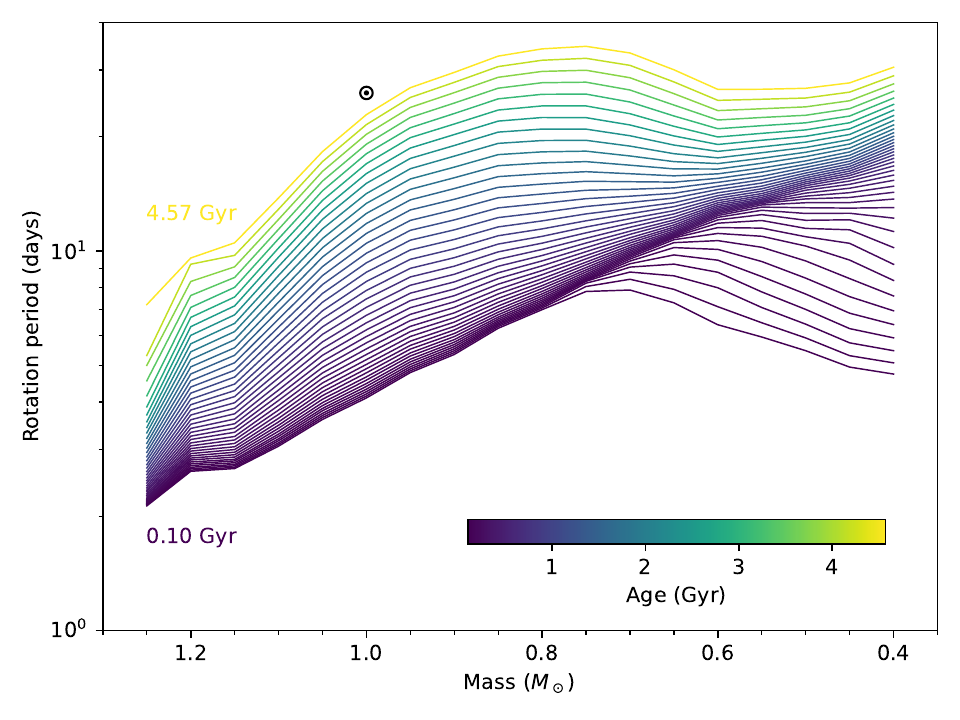}
\caption{Evolution of the \srs{} according to model ``bpl'': rotational isochrones from $0.1$ to $4.57$ Gyr.}
\label{fig:piso_sbpl2}
\end{center}
\end{figure}

Figure~\ref{fig:piso_sbpl2} shows isochrones from $0.1$ to $4.57$ Gyr in the usual period vs. mass diagram. 
A piling up of the isochrones of age up to $\approx 2$ Gyr is clearly visible, with stars of lower mass converging onto the ridge with increasingly longer timescale.
This over-density of isochrones corresponds to a period of slowed down rotational evolution, and is a direct consequence of the stalled spin-down phase due to the resurfacing of angular momentum from the interior via the (mass-dependent) re-coupling mechanism.

\begin{figure}[h!]
\begin{center}
\includegraphics[width=0.49\textwidth, trim=1.3cm 0.5cm 1.5cm 0.5cm, clip]{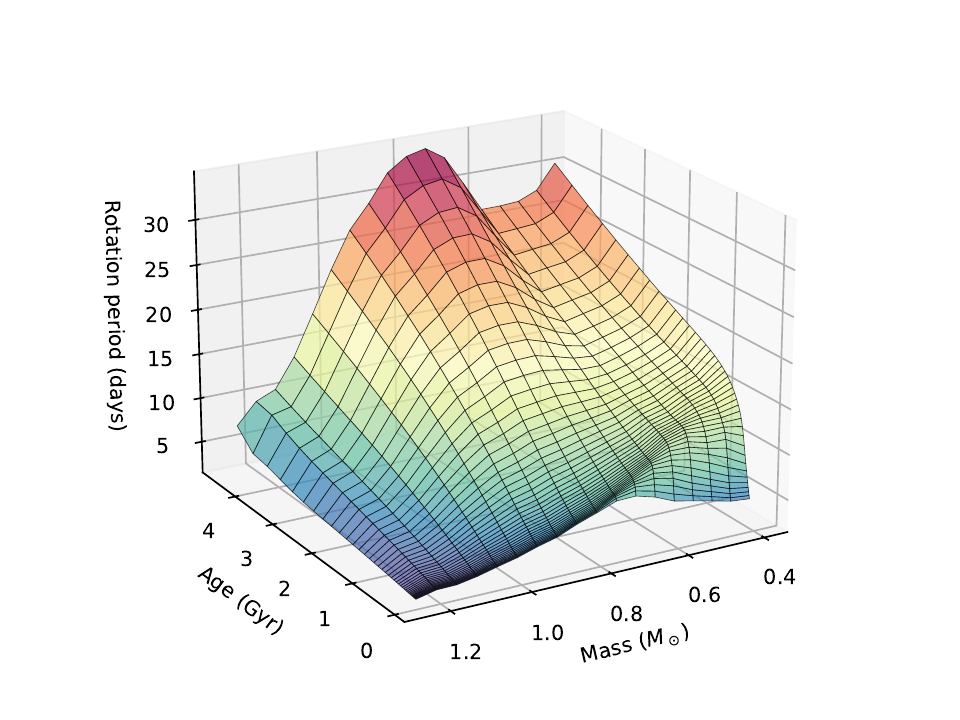}
\caption{Evolution of the \srs{} according to model ``bpl'': three-dimensional representation of the mass--age--rotation period relation.}
\label{fig:piso_sbpl3}
\end{center}
\end{figure}

Figure~\ref{fig:piso_sbpl3} shows a three-dimensional representation of the full mass--age--rotation period relation.
This view illustrates how the rotation period depends on both stellar mass and age. 
It is clear, in particular, that the rotation period cannot be adequately represented by a functional relation in which the dependences on mass and age enter as separate, independent factors.

Table~\ref{tab:piso_sbpl} lists the surface rotation period as a function of age, with each column corresponding to a different stellar mass. 
In other words, each row is an isochrone, whose age is given by the first entry in the row. 
The de-reddened Gaia $(G_{\rm BP}-G_{\rm RP})_0$ color corresponding to each mass is also given in the bottom row for reference.
For the purposes of the Table, the mass-to-color conversion has been performed using a MIST isochrone of solar metallicity and age equal to $1$ Gyr.

\begin{table*}[h]
\caption{Rotational isochrones (rows)  and rotational histories (columns) for stars on the \srs{} according to the ``bpl'' model. Units of age, mass, and rotation period are Gyr, $M_\odot$, and days, respectively.}
\begin{center}
\rotatebox{90}{
\begin{tabular}{c | ccc ccc ccc ccc ccc ccc}
\hline
\hline
Age & \multicolumn{18}{c}{Mass} \\
\hline
      &  0.40 &  0.45 &  0.50 &  0.55 &  0.60 &  0.65 &  0.70 &  0.75 &  0.80 &  0.85 &  0.90 &  0.95 &  1.00 &  1.05 &  1.10 &  1.15 &  1.20 &  1.25 \\
      \hline
0.10 &  4.74 &  4.95 &  5.46 &  5.93 &  6.39 &  7.30 &  7.89 &  7.82 &  7.01 &  6.26 &  5.34 &  4.78 &  4.10 &  3.60 &  3.06 &  2.67 &  2.63 &  2.13 \\
0.11 &  5.20 &  5.43 &  6.06 &  6.67 &  7.42 &  8.14 &  8.56 &  8.13 &  7.10 &  6.34 &  5.42 &  4.85 &  4.17 &  3.66 &  3.10 &  2.70 &  2.65 &  2.14 \\
0.13 &  5.75 &  6.06 &  6.81 &  7.56 &  8.54 &  9.04 &  8.99 &  8.30 &  7.19 &  6.42 &  5.50 &  4.93 &  4.24 &  3.72 &  3.15 &  2.73 &  2.69 &  2.15 \\
0.15 &  6.40 &  6.86 &  7.70 &  8.56 &  9.46 &  9.77 &  9.26 &  8.37 &  7.27 &  6.49 &  5.60 &  5.03 &  4.33 &  3.80 &  3.20 &  2.77 &  2.72 &  2.16 \\
0.17 &  7.17 &  7.85 &  8.78 &  9.67 & 10.29 & 10.33 &  9.45 &  8.42 &  7.37 &  6.57 &  5.70 &  5.13 &  4.43 &  3.87 &  3.26 &  2.81 &  2.76 &  2.17 \\
0.19 &  8.10 &  9.12 & 10.04 & 10.72 & 10.97 & 10.66 &  9.60 &  8.50 &  7.48 &  6.67 &  5.81 &  5.26 &  4.54 &  3.96 &  3.33 &  2.86 &  2.80 &  2.18 \\
0.21 &  9.22 & 10.48 & 10.91 & 11.49 & 11.53 & 10.83 &  9.72 &  8.61 &  7.61 &  6.79 &  5.95 &  5.39 &  4.66 &  4.07 &  3.41 &  2.91 &  2.85 &  2.20 \\
0.24 & 10.56 & 11.64 & 11.67 & 12.19 & 11.93 & 10.92 &  9.84 &  8.73 &  7.75 &  6.94 &  6.10 &  5.55 &  4.81 &  4.19 &  3.50 &  2.97 &  2.90 &  2.22 \\
0.28 & 11.91 & 12.43 & 12.40 & 12.70 & 12.21 & 11.01 &  9.96 &  8.88 &  7.92 &  7.12 &  6.29 &  5.74 &  4.98 &  4.33 &  3.60 &  3.04 &  2.96 &  2.24 \\
0.31 & 13.02 & 12.97 & 13.00 & 13.05 & 12.40 & 11.11 & 10.10 &  9.05 &  8.11 &  7.33 &  6.50 &  5.96 &  5.18 &  4.49 &  3.71 &  3.11 &  3.02 &  2.27 \\
0.36 & 13.91 & 13.56 & 13.52 & 13.27 & 12.56 & 11.23 & 10.26 &  9.25 &  8.34 &  7.57 &  6.74 &  6.20 &  5.40 &  4.67 &  3.85 &  3.20 &  3.09 &  2.30 \\
0.41 & 14.63 & 14.11 & 13.88 & 13.41 & 12.70 & 11.39 & 10.45 &  9.47 &  8.59 &  7.85 &  7.02 &  6.49 &  5.66 &  4.87 &  3.99 &  3.30 &  3.17 &  2.34 \\
0.46 & 15.35 & 14.64 & 14.21 & 13.51 & 12.85 & 11.56 & 10.65 &  9.73 &  8.88 &  8.18 &  7.35 &  6.81 &  5.95 &  5.12 &  4.17 &  3.41 &  3.25 &  2.38 \\
0.52 & 16.00 & 15.09 & 14.53 & 13.63 & 13.01 & 11.77 & 10.91 & 10.05 &  9.23 &  8.56 &  7.73 &  7.19 &  6.29 &  5.38 &  4.36 &  3.54 &  3.36 &  2.42 \\
0.60 & 16.61 & 15.48 & 14.80 & 13.80 & 13.19 & 12.01 & 11.20 & 10.39 &  9.63 &  9.00 &  8.18 &  7.62 &  6.68 &  5.69 &  4.57 &  3.69 &  3.47 &  2.47 \\
0.68 & 17.16 & 15.80 & 15.04 & 14.00 & 13.40 & 12.28 & 11.55 & 10.81 & 10.10 &  9.51 &  8.68 &  8.13 &  7.11 &  6.03 &  4.81 &  3.85 &  3.59 &  2.53 \\
0.77 & 17.66 & 16.09 & 15.28 & 14.23 & 13.64 & 12.60 & 11.94 & 11.29 & 10.64 & 10.10 &  9.27 &  8.69 &  7.60 &  6.42 &  5.09 &  4.03 &  3.74 &  2.59 \\
0.87 & 18.14 & 16.39 & 15.53 & 14.51 & 13.92 & 12.98 & 12.40 & 11.86 & 11.27 & 10.79 &  9.95 &  9.34 &  8.17 &  6.86 &  5.39 &  4.23 &  3.89 &  2.66 \\
0.99 & 18.59 & 16.70 & 15.81 & 14.83 & 14.23 & 13.40 & 12.95 & 12.53 & 12.01 & 11.58 & 10.72 & 10.07 &  8.80 &  7.33 &  5.73 &  4.46 &  4.07 &  2.74 \\
1.13 & 19.06 & 17.05 & 16.13 & 15.21 & 14.62 & 13.91 & 13.59 & 13.30 & 12.86 & 12.49 & 11.59 & 10.89 &  9.49 &  7.88 &  6.11 &  4.72 &  4.27 &  2.82 \\
1.28 & 19.54 & 17.44 & 16.50 & 15.62 & 15.04 & 14.51 & 14.35 & 14.21 & 13.86 & 13.54 & 12.59 & 11.81 & 10.26 &  8.46 &  6.51 &  5.00 &  4.49 &  2.92 \\
1.45 & 20.06 & 17.87 & 16.92 & 16.14 & 15.54 & 15.19 & 15.23 & 15.27 & 15.01 & 14.72 & 13.69 & 12.83 & 11.13 &  9.10 &  6.97 &  5.31 &  4.74 &  3.03 \\
1.65 & 20.64 & 18.38 & 17.42 & 16.70 & 16.13 & 15.99 & 16.26 & 16.50 & 16.35 & 16.06 & 14.93 & 13.95 & 12.05 &  9.83 &  7.48 &  5.66 &  5.03 &  3.15 \\
1.87 & 21.31 & 18.98 & 18.05 & 17.37 & 16.80 & 16.93 & 17.48 & 17.92 & 17.86 & 17.55 & 16.30 & 15.16 & 13.07 & 10.59 &  8.02 &  6.03 &  5.35 &  3.28 \\
2.13 & 22.07 & 19.65 & 18.74 & 18.16 & 17.60 & 18.05 & 18.90 & 19.57 & 19.55 & 19.21 & 17.78 & 16.48 & 14.17 & 11.44 &  8.63 &  6.45 &  5.70 &  3.43 \\
2.42 & 22.98 & 20.51 & 19.58 & 19.05 & 18.55 & 19.36 & 20.55 & 21.45 & 21.47 & 21.02 & 19.41 & 17.92 & 15.38 & 12.36 &  9.28 &  6.91 &  6.10 &  3.60 \\
2.75 & 24.05 & 21.49 & 20.58 & 20.12 & 19.67 & 20.89 & 22.48 & 23.55 & 23.56 & 23.01 & 21.16 & 19.47 & 16.65 & 13.35 &  9.99 &  7.43 &  6.55 &  3.81 \\
3.12 & 25.29 & 22.68 & 21.77 & 21.39 & 21.00 & 22.69 & 24.69 & 25.91 & 25.88 & 25.17 & 23.03 & 21.13 & 18.05 & 14.41 & 10.78 &  8.02 &  7.10 &  4.15 \\
3.54 & 26.74 & 24.09 & 23.16 & 22.88 & 22.57 & 24.79 & 27.22 & 28.55 & 28.38 & 27.48 & 25.06 & 22.95 & 19.51 & 15.59 & 11.65 &  8.70 &  7.83 &  4.72 \\
4.02 & 28.47 & 25.75 & 24.83 & 24.63 & 24.43 & 27.22 & 30.08 & 31.44 & 31.14 & 29.93 & 27.24 & 24.88 & 21.14 & 16.87 & 12.63 &  9.51 &  8.91 &  5.00 \\
4.57 & 30.49 & 27.73 & 26.84 & 26.68 & 26.65 & 30.03 & 33.25 & 34.62 & 34.09 & 32.62 & 29.56 & 26.94 & 22.88 & 18.28 & 13.80 & 10.51 &  9.58 &  7.22 \\
\hline
     &  2.08 &  1.98 &  1.88 &  1.78 &  1.66 &  1.51 &  1.36 &  1.24 &  1.13 &  1.02 &  0.93 &  0.87 &  0.80 &  0.75 &  0.71 &  0.66 &  0.62 &  0.58 \\
\hline
Age & \multicolumn{18}{c}{$(G_{\rm BP}-G_{\rm RP})_0$} \\
\hline
\end{tabular}
}
\end{center}
\label{tab:piso_sbpl}
\end{table*}

\end{document}